\documentclass[%
 reprint,
 amsmath,amssymb,
 aps,
]{revtex4-2}

\usepackage{graphicx}
\usepackage{dcolumn}
\usepackage{bm}
\usepackage{color}

\begin{document}

\preprint{APS/123-QED}

\title{Function Smoothing Regularization for Precision Factorization Machine Annealing in Continuous Variable Optimization Problems}

\author{Katsuhiro Endo}
 \email{katsuhiro.endo@aist.go.jp}
\author{Kazuaki Z. Takahashi}%
 \email{kazu.takahashi@aist.go.jp}
\affiliation{%
 National Institute of Advanced Industrial Science and Technology (AIST),\\
 Research Center for Computational Design of Advanced Functional Materials,\\
 Central 2, 1-1-1 Umezono, Tsukuba, Ibaraki 305-8568, Japan
}%



\date{\today}

\begin{abstract}
Solving continuous variable optimization problems by factorization machine quantum annealing (FMQA) demonstrates the potential of Ising machines to be extended as a solver for integer and real optimization problems.
However, the details of the Hamiltonian function surface obtained by factorization machine (FM) have been overlooked.
This study shows that in the widely common case where real numbers are represented by a combination of binary variables, the function surface of the Hamiltonian obtained by FM can be very noisy.
This noise interferes with the inherent capabilities of quantum annealing and is likely to be a substantial cause of problems previously considered unsolvable due to the limitations of FMQA performance.
The origin of the noise is identified and a simple, general method is proposed to prevent its occurrence.
The generalization performance of the proposed method and its ability to solve practical problems is demonstrated.
\end{abstract}

\maketitle


\section{\label{sec:intro}Introduction}
Ising machines, including quantum annealing machines, are a type of special-purpose computer for solving combinatorial optimization problems in an approximate and efficient manner\cite{tiunov2019annealing,king2022coherent,mohseni2022ising,izawa2022continuous}.
Since problems such as path planning, scheduling, and layout optimization can be formulated as combinatorial optimization problems, the use of Ising machines has been increasing in recent years\cite{rieffel2015case,carugno2022evaluating,klar2022quantum}.
By inputting an energy function defined by a quadratic form of binary variables, called a quadratic unconstrained binary optimization problem (QUBO), Ising machines are expected to obtain a combination of variables that minimizes the energy function\cite{glover2018tutorial}.
The energy function, the Hamiltonian, is an objective function that should be designed according to the combinatorial optimization problem to be solved.
However, in most cases, the Hamiltonian is not given analytically for combinatorial optimization problems using data obtained from actual experiments and simulations.
Even in such cases, the evaluated value of the objective function at a certain combination of optimization variables (hereinafter referred to as the evaluation point) can be obtained from experiments or simulations.
Therefore, it is possible in principle to learn the Hamiltonian by exploratory evaluation of various evaluation points in the manner of black box optimization\cite{guidotti2018survey,hansen2010comparing}.
Factorization machine quantum annealing (FMQA)\cite{kitai2020designing} is a form of black-box optimization using Ising machines and is a combination of quantum annealing (QA) and factorization machine (FM), a type of machine learning method.
In FMQA, FM learns an unknown objective function with binary variables as optimization variables as a QUBO-style Hamiltonian, and then performs quantum annealing based on the obtained Hamiltonian to obtain the next evaluation point.
Kitani et al. demonstrated that FMQA can design thermo-functional materials with fewer evaluation points than Gaussian process regression, but with comparable quality\cite{kitai2020designing}.
FMQA, including examples where QA is substituted for classical Ising machines, has been widely used in the design of antimicrobial peptides\cite{tucs2023quantum}, material design of spintronics devices\cite{nawa2023quantum}, resonance avoidance structure design of printed circuit boards\cite{matsumori2022application}, molecular design for highly efficient OLEDs\cite{gao2023quantum}, spatial distribution optimization of photon crystal surface-issued lasers\cite{inoue2022towards}, proposal of a recommendation system\cite{liu2022implementation}, and so on.

As mentioned above, FMQA uses pairs of binary variables as optimization variables, but it is also possible to embed continuous variables such as integers and real numbers into binary variables.
With this embedding, FMQA is expected to be extended as a solver for integer and real number optimization problems.
In FMQA, methods have been proposed to compress continuous variables (including integers and real numbers) into binary variables using a binary variational autoencoder\cite{mao2023chemical} or to represent continuous variables as a combination of binary variables\cite{wilson2021machine}.
However, these methods do not consider the details of the energy function surface obtained by FM.

In this study, we show that in the widely common case of representing real numbers by a combination of binary variables, the function surface of the Hamiltonian obtained by FM can be extremely noisy.
This noise interferes with the inherent ability of QA to solve black-box optimization and may be a substantial cause of problems previously considered unsolvable due to limitations in FMQA performance.
We identify the origin of noise and propose a simple, general method to prevent its occurrence.

\section{Methodology}
\subsection{Factorization Machine Quantum Annealing}
The Ising machine attempts to minimize the Hamiltonian $H_{\rm QUBO}$, expressed as
\begin{eqnarray}
H_{\rm QUBO}(\{ x_{i} \}) = \sum_{i,j=1}^{N} Q_{ij} x_{i} x_{j},
\label{eq:h_qubo}
\end{eqnarray}
where the brackets $\{\cdots\}$ denote a set, $\{ x_i \}$ is the set of $N$ binary variables to be optimized and $Q_{ij}$ is the real coefficient matrix that determines the shape of the Hamiltonian.
A Hamiltonian with such a set of binary variables as the optimization variables is called a Hamiltonian of the QUBO form.
The Ising machine is a computer that receives an input of $Q_{ij}$ and outputs $\{ x_i \}$ that minimizes $H_{\rm QUBO}$.
Note again that in most cases, the Hamiltonian is not given analytically in combinatorial optimization problems with data obtained from actual experiments or simulations.
Here we define $H$ as the true Hamiltonian corresponding to the data obtained from experiments and simulations, which is clearly distinguished from $H_{\rm QUBO}$.

\begin{figure}
\includegraphics[width=0.4\textwidth]{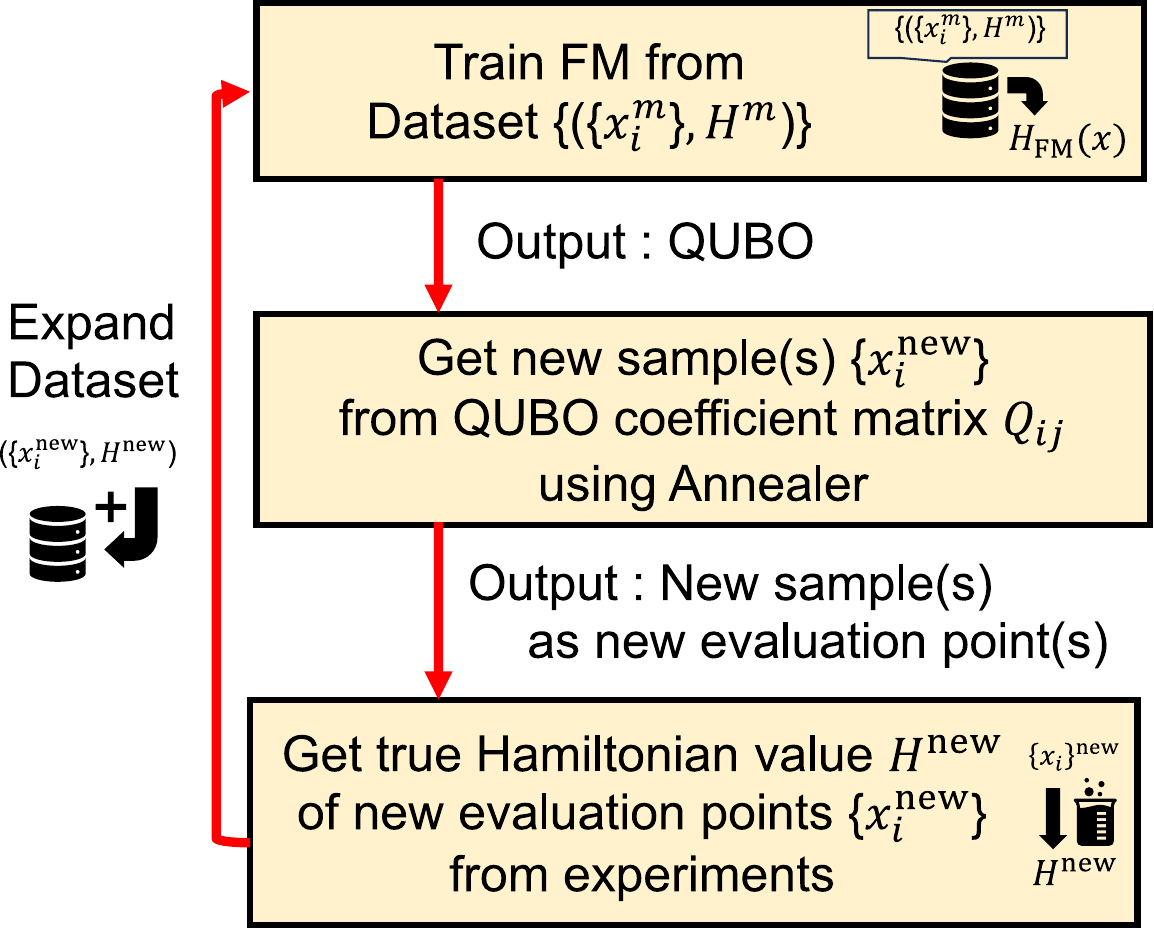}
\caption{
FMQA flow.
}
\label{fig:flowchart}
\end{figure}

FMQA is a black-box optimization algorithm using Ising machines, as shown in Figure~\ref{fig:flowchart}.
First, a known evaluation point $\{ {x_i}^{m} \}$ and its corresponding true Hamiltonian ${H}^{m}$ (evaluation value) pair $( \{ {x_i}^{m} \},{H}^{m} )$ are used as training data.
Here $m$ is the serial number of the training data, and the dataset is $\{ ( \{ {x_i}^{m} \},{H}^{m} ) \}$ (see Fig.~\ref{fig:flowchart}).
FM uses the dataset to estimate the Hamiltonian in QUBO format.
The FM models the Hamiltonian as
\begin{eqnarray}
H_{\rm FM}(\{ x_{i} \}) = a + \sum_{i=1}^{N}b_{i}x_{i} + \sum_{i=j=1}^{N} \langle {\boldsymbol{v}_{i}} \cdot {\boldsymbol{v}_{j}} \rangle x_{i} x_{j},
\label{eq:h_fm}
\end{eqnarray}
where $a$ and $b_i$ are FM model parameters for real scalar and $\boldsymbol{v}_{i}$ is the FM model parameter for $K$-dimensional real vector.
Eq.~\ref{eq:h_fm} is in QUBO form except for $a$, given that $x_{i}^{2} \equiv x_{i}$.
The $a$ can be ignored since it is irrelevant to the optimization.
The situation where the data handled by FM is sparse can easily be assumed.
Since FM can reduce the expressive power of the model by appropriately restricting the dimension $K$ of $\boldsymbol{v}_{i}$, high performance can be achieved by preventing overfitting even with sparse data.
We refer to $K$ as the rank of the FM.

Training by FM aims to minimize the following loss function $L$:
\begin{eqnarray}
L(a,\{ b_{i} \},\{ \boldsymbol{v}_{i} \}) = \sum_{m} \left( H_{\rm FM} (\{ x_{i}^{m} \}) - H^{m} \right)^{2}.
\end{eqnarray}
Since $L$ is differentiable with respect to the parameters $a$,
$\{ b_{i} \}$, and $\{ \boldsymbol{v}_{i} \}$, the optimal parameter values can be trained by general gradient descent methods, including stochastic gradient descent methods.
In this work, AMSGRAD\cite{tan2019convergence}, a type of gradient descent method, is employed.

After training $H_{\rm FM}$ from the dataset, the new evaluation point $\{ x_i^{\rm new} \}$ is obtained by inputting the QUBO coefficient matrix $Q_{ij}$ into the Ising machine (see Fig.~\ref{fig:flowchart}).
Note that if $\{ x_i^{\rm new} \}$ is not included in the known evaluation points, the $H_{\rm FM}$ corresponding to $\{ x_i^{\rm new} \}$ does not coincide with the $H$ corresponding to $\{ x_i^{\rm new} \}$ (=$H^{\rm new}$).
${H}^{\rm new}$ is then obtained by experiments or simulations and added to the dataset (see Fig.~\ref{fig:flowchart}).
FMQA is a black-box optimization algorithm that efficiently searches for the true optimal solution by repeating the aforementioned three processes: training $H_{\rm FM}$ using FM, sampling the new evaluation point using the Ising machine, and obtaining the new evaluation value from experiments/simulations.
Hereafter, the three processes shown in Fig.~\ref{fig:flowchart} will be referred to as one step.

\subsection{One-hot representation for continuous values}
The original FMQA performs black-box optimization for problems in which binary variables are the optimization variables.
However, Seki et al. have succeeded in performing black-box optimization for continuous values by embedding integer continuous values in an array of binary variables\cite{seki2022black}.
One-hot representation is one of their proposed embedding methods, where the continuous value is represented as the array of binary variables as follows:
\begin{eqnarray}
 y(c) &\rightarrow& [ x_{1}, \cdots x_{i} \cdots x_{C} ]_{\rm one-hot}, \nonumber \\
 &&x_{i} =
 \begin{cases}
  & 1 \quad i = c \\
  & 0 \quad i \neq c,
 \end{cases}
\label{eq:one_hot}
\end{eqnarray}
where $y(c)$ is the $c$-th value of the $C$ (= $N$) continuous value $y$, the arrow means encoding to the one-hot representation, and the brackets $[\cdots]_{\rm one-hot}$ denote the one-hot vector.
Obviously, the one-hot representation has the restriction that only one of the $C$ binary variables is 1 and the others are 0.
However, the Ising machine cannot handle this constraint directly.
Therefore, the following correction term is added to $H_{\rm FM}$ that imposes an energy penalty when the one-hot representation is not satisfied:
\begin{eqnarray}
H_{\rm one-hot} = \alpha_{\rm one-hot} \left [ \left( \sum_{k=1}^{C} x_{k} \right ) - 1 \right ]^{2},
\label{eq:penalty}
\end{eqnarray}
where $\alpha_{\rm one-hot}$ is the strength of the penalty.
Eq.~\ref{eq:penalty} takes zero only when the one-hot representation is satisfied; otherwise it takes positive values.
Thus, when $\alpha_{\rm one-hot}$ is sufficiently large, the one-hot representation is naturally satisfied by Hamiltonian minimization with the Ising machine.
Note that the $Q_{ij}$ input to the Ising machine corresponds to $H_{\rm FM} + H_{\rm one-hot}$.

When there are multiple continuous values to be optimized, different binary variables are assigned to each continuous value.
For example, two $C$ (= $N/2$) continuous values $y_1$ and $y_2$ are represented as follows:
\begin{eqnarray}
 y_{1}(c) &\rightarrow& [ x_{1}, \cdots x_{i} \cdots x_{C} ]_{\rm one-hot}, \nonumber \\
 &&x_{i} =
 \begin{cases}
  & 1 \quad i = c \\
  & 0 \quad i \neq c,
 \end{cases} \\
 y_{2}(c) &\rightarrow& [ x_{C+1}, \cdots x_{i} \cdots x_{2C} ]_{\rm one-hot}, \nonumber \\
 &&x_{i} =
 \begin{cases}
  & 1 \quad i = c \\
  & 0 \quad i \neq c.
 \end{cases}
\label{eq:one_hot_2}
\end{eqnarray}
$H_{\rm one-hot}$ is also added by the number of corresponding continuous values.

\subsection{Continuous Variable Optimization Problems}
In order to carefully investigate the resulting energy function surfaces of the FMQA scheme in continuous variable optimization problems, this study addresses three problems presented in the following subsections.

\subsubsection{Simple Toy Hamiltonian}
The simplest problem setup is a toy system where the true Hamiltonian is analytic and there is no interaction between continuous variables.
We define the following true Hamiltonian with only two continuous variables,
\begin{eqnarray}
H_{1} = y_{1}^{2} + 2y_{2}^{2},
\label{eq:simple_toy}
\end{eqnarray}
where $y_1$ and $y_2$ are both 101 real continuous values equally divided into 100 over the interval [-5.12, 5.12].
The solution that minimizes this Hamiltonian is clearly $y_{1}=y_{2}=0$ and is contained in the values that $y_1$ and $y_2$ can take.
This simple problem setup aids in understanding the results of the FMQA scheme because it is easy to compare the function surface given by $H_{\rm FM}$ with that given by $H_1$.

\subsubsection{Interactive Toy Hamiltonian}
A simple but problem that allows evaluation of FM's generalization performance is a toy system in which the true Hamiltonian is analytic, but continuous variables interact with each other.
The true Hamiltonian with four continuous variables is defined by the following equation, 
\begin{eqnarray}
H_{2} = (y_{1}^{2} + y_{2}^{2} + y_{3}^{2} + y_{4}^{2})^{1/2},
\label{eq:interactive_toy}
\end{eqnarray}
where $y_{1-4}$ are 101 real continuous values equally divided into 100 over the interval [-1, 1].
Unlike the continuous variables in $H_1$, those in $H_2$ are interactive.
In other words, the $H_2$ change for a change in one continuous variable depends on the values of the other continuous variables.

\subsubsection{Hamiltonian to predict nanophysical properties of contrast agents}
Lipid-coated microbubbles (LCMB) are a typical medical contrast agent that has been commercialized, applying the difference in acoustic impedance between biological tissue and bubbles\cite{VERSLUIS20202117,FRINKING2020892}.
The following model equations that precisely predicts LCMB oscillations have been proposed by Marmottant et al.\cite{marmottant2005model},
\begin{eqnarray}
\rho_{\rm L} &&\left ( R \frac{d^{2} R}{d^{2} t} + \frac{3}{2} \frac{dR}{dt} \right ) = \nonumber \\
&&\left ( P_{0} + \frac{2 \sigma_{0}}{R_0} \right ) \left ( \frac{R_0}{R} \right )^{3 \kappa} \left ( 1-\frac{3 \kappa}{c_{\rm L}} \frac{dR}{dt} \right ) \nonumber \\
&&- P_{0} - P_{\rm A}(t) - \frac{4 \mu}{R} \frac{dR}{dt} - \frac{\sigma (R)}{R} - \frac{4 \kappa_{S}}{R^2} \frac{dR}{dt}, \\
\sigma (R) &&=
 \begin{cases}
  & \chi \left ( \frac{R^2}{R_0^2} - 1 \right ) \quad R > R_{\rm b} \\
  & 0 \qquad \qquad \qquad \! \! R < R_{\rm b}, \\
 \end{cases} \\
R_{\rm b} &&= R_{0}[1 + (\sigma_{0}/\chi)]^{-1/2},
\label{eq:marmottant}
\end{eqnarray}
where $\rho_{\rm L}$ is the liquid density, $R$ is the bubble radius, $t$ is the time, $\kappa$ is the polytropic gas exponent, $c_{\rm L}$ is the speed of sound in liquid, $\mu$ is the liquid viscosity, $\kappa_{\rm S}$ is the surface dilatational viscosity, $P_0$ is the ambient pressure, $R_0$ is the equilibrium radius of bubble, $P_{\rm A}(t)$ is the acoustic pressure, $\sigma (R)$ is the effective surface tension, $\sigma_0$ is the natural surface tension, $R_{\rm b}$ is the buckling radius, and $\chi$ is the shell elastic compression modulus.
Clearly, the equation is of Rayleigh-Plesset type.

The three key nanophysical parameters of the model are $\chi$, $\kappa_{\rm S}$, and $\sigma_0$, and are difficult to measure experimentally, but can be estimated by fitting model oscillations to experimental results\cite{jafari2021toward}.
In this study, we propose an alternative to this fitting process using FMQA.
That is, the true Hamiltonian is expressed by the following equation,
\begin{eqnarray}
H_{3} = \int_{t} dt[R(t; \chi, \kappa_{\rm S}, \sigma_0) - R^{*}(t)]^{2},
\label{eq:hamitonian_lcbm}
\end{eqnarray}
where $R(t; \chi, \kappa_{\rm S}, \sigma_0)$ means that all parameters that determine the time evolution of $R$ in the model equations except $\chi$, $\kappa_{\rm S}$, and $\sigma_0$ are fixed, and $R^{*}(t)$ means the true time evolution of $R$.
Note that an analytical representation of the true Hamiltonian is difficult because it is based on nonlinear oscillations that encompass three parameters.

We approximate $H_3$ by discretizing time.
First, $R^{*}(t)$ ($0 \leq t \leq 10\,\mu {\rm s}$), obtained from Ref.~\citenum{overvelde2010nonlinear}, is discretized every 0.005\,$\mu$s into a sequence of 201 real numbers.
Here, $R^{*}(t)$ corresponds to the Marmottant and co-worker's model with $\rho_{\rm L}=10^3$\,kg/l, $\kappa=1.07$, $c_{\rm L}=$\,1500\,m/s, $\mu =10^{-3}$\,Pa$\cdot$s, $\kappa_{\rm S}=6.0 \times 10^{-9}$\,kg/s, $P_{0}=10^5$\,Pa, $R_{0}=3.2 \times 10^{-6}$\,m, $\sigma_{0}=0.02$\,N/m, $\chi=2.5$\,N/m, and $P_{\rm A}(t)$ is the Gaussian-tapered sound pressure of 25\,kPa and 1.5\,MHz\cite{overvelde2010nonlinear}.

Next, the data for $R(t; \chi, \kappa_{\rm S}, \sigma_0)$ are generated from the Marmottant and co-worker's model by fixing parameters other than $\chi$, $\kappa_{\rm S}$, and $\sigma_0$ based on Ref.~\citenum{overvelde2010nonlinear}.
The $\chi$, $\kappa_{\rm S}$, and $\sigma_0$ are assumed to be unknown and are represented by 65 continuous values equally divided into 64 over the intervals [1.0\,N/m,4.0\,N/m], [0,12.0$\times 10^{-9}$\,kg/s], and [0.01\,N/m,0.03\,N/m], respectively.
Note that $R(t; \chi, \kappa_{\rm S}, \sigma_0)$ is discrete with respect to time, as is $R^{*}(t)$.

Finally, $H_3$ is computed for the discretized time to obtain an approximation of the true Hamiltonian.

\section{Results and Discussion}
\subsection{Noisy energy surface and the function smoothing regularization to resolve it}
Let us first observe the functional surfaces acquired by the naive FM for the simple toy Hamiltonian $H_1$.
The black box optimization was performed according to the flow shown in Fig.~\ref{fig:flowchart}.
As the first dataset, 16 evaluation points and their corresponding true Hamiltonian values were selected from all combinations of $y_1$ and $y_2$ (101 $\times$ 101 = 10201 possible combinations) using uniform random numbers.
The learning rate for AMSGRAD was set at 0.1.
Although Ising machines or quantum annealing can be used to sample the next evaluation point from the learned Hamiltonian, here we substituted sampling from the following Boltzmann distribution $p$ to eliminate the influence of the device,
\begin{eqnarray}
p(y_{1},y_{2}) = Z^{-1} \exp(- \beta H_{\rm FM}),
\label{eq:boltzmann}
\end{eqnarray}
where $Z$ is the normalization constant and $\beta = 20.0$ is the inverse temperature is the inverse temperature.
After 16 steps, FM was terminated when the number of known evaluation points reached 272.

\begin{figure}
\includegraphics[width=0.48\textwidth]{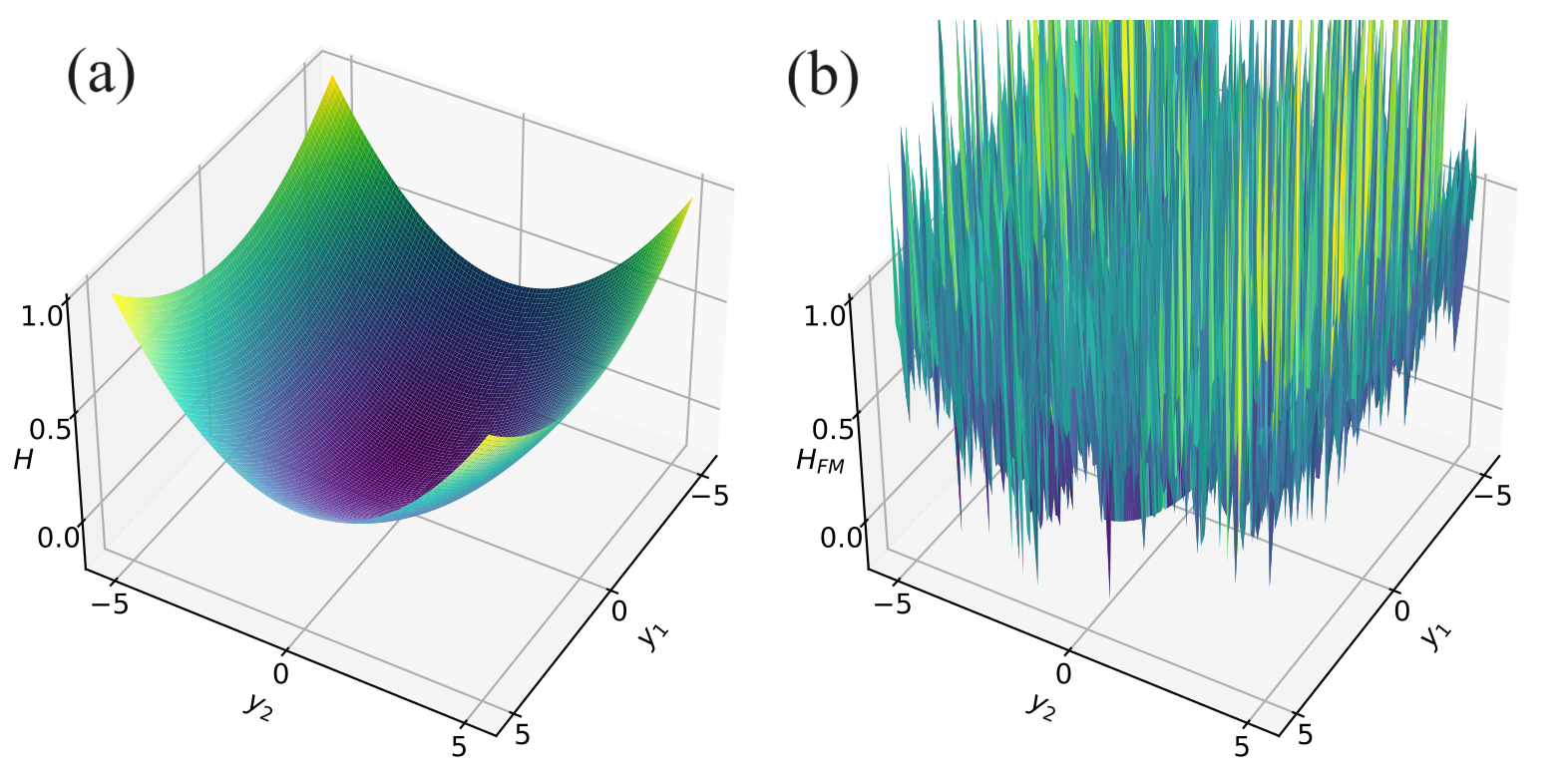}
\caption{
Function surfaces of (a) $H_1$, and (b) $H_{\rm FM}$ after 16 steps by naive FM.
}
\label{fig:h1_surface}
\end{figure}

Figure~\ref{fig:h1_surface}(a) shows the function surface of $H_1$, and (b) shows the function surface of $H_{\rm FM}$ with 16 steps of training by naive FM.
Even though $H_1$ has a very simple function surface, the function surface acquired by naive FM was very noisy, suggesting that it may be difficult to suggest the next evaluation point from this function surface.

\begin{figure}
\includegraphics[width=0.48\textwidth]{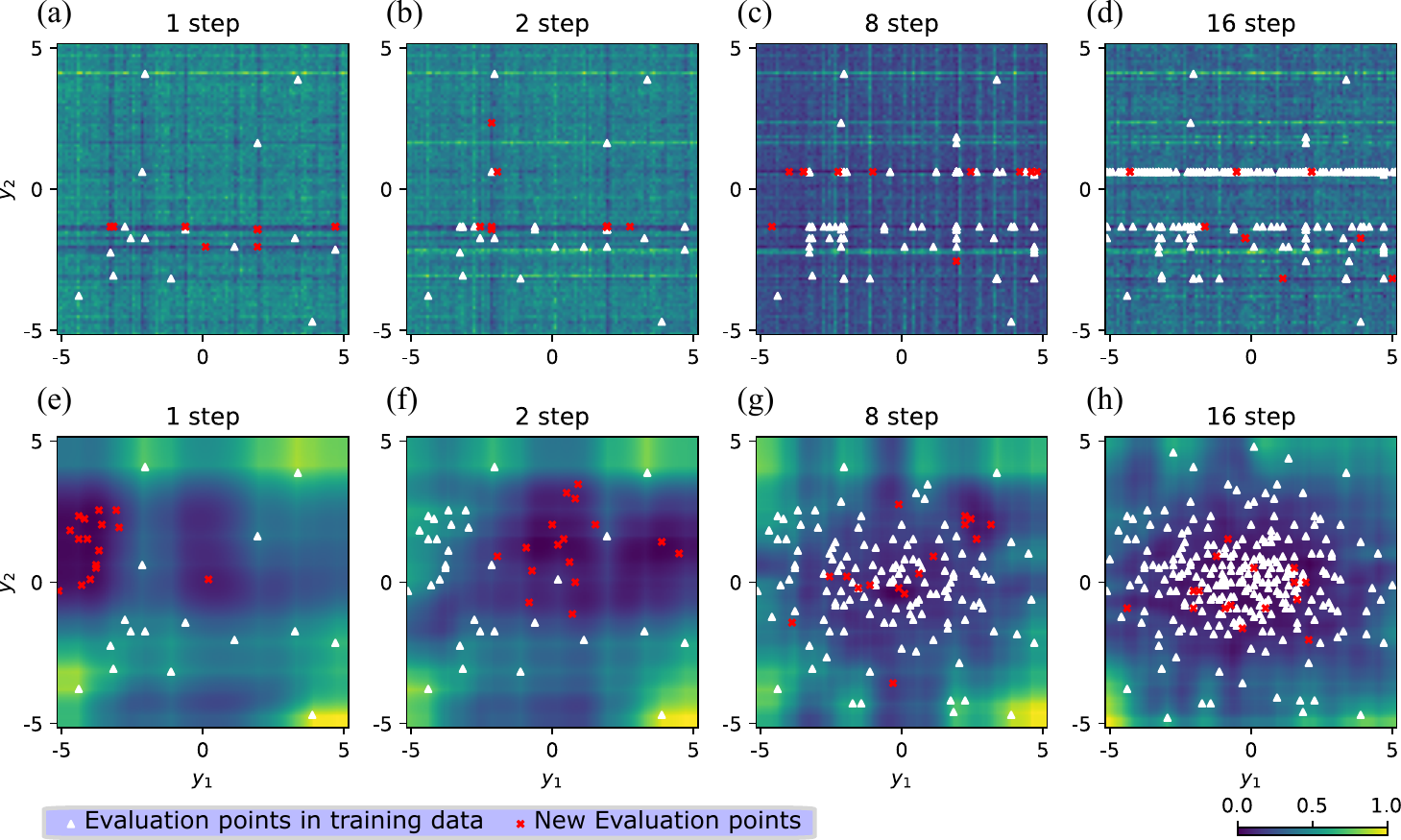}
\caption{
Function surfaces obtained by (a)-(d) naive FM immediately after steps 1, 2, 8, and 16, respectively, and by (e)-(h) FSRFM immediately after steps 1, 2, 8, and 16, respectively.
}
\label{fig:h1_compare}
\end{figure}

Figure~\ref{fig:h1_compare}(a), (b), (c), and (d) show the function surfaces obtained by naive FM immediately after steps 1, 2, 8, and 16, respectively.
Each white point is $(y_1^m, y_2^m)$ and each red point is $(y_1^{\rm new}, y_2^{\rm new})$, each sampled from Eq.~\ref{eq:boltzmann}.
At every step, the red points were concentrated in regions where the value of $H_{\rm FM}$ was relatively small.
It is clear from Fig.~\ref{fig:h1_compare}(d) that $H_{\rm FM}$ did not approximate $H_1$ at all after 16 steps.
The above means that extreme bias in sampling of $(y_1^{\rm new}, y_2^{\rm new})$ occurred and the black box optimization did not work well.
Note here that the training FM itself is well underway, since FM is able to predict the Hamiltonian almost correctly only for the samples included in the dataset.

In order to clarify the cause of the naive FM acquiring the very noisy function surface, we focused on the minimization process of the loss function $L$ in the FM.
Given the dataset $\{ (\{ x_{i}^{m} \}, H^{m}) \}$ used to train the FM, the FM is updated using the gradients of $L$ shown in the following equations,
\begin{eqnarray}
&&\frac{\partial L}{\partial b_{l}} = \frac{\partial}{\partial b_{l}} \sum_{m} \left ( H_{\rm FM} (\{ x_{i}^{m} \}) - H^{m} \right )^{2} \nonumber \\
&&= \sum_{m} 2 \left ( H_{\rm FM} (\{ x_{i}^{m} \}) - H^{m} \right ) \frac{\partial}{\partial b_{l}} H_{\rm FM}(\{ x_{i}^{m} \}) \nonumber \\
&&= \sum_{m} 2 \left ( H_{\rm FM} (\{ x_{i}^{m} \}) - H^{m} \right ) x_{l}^{m} \label{eq:gradient_b}\\
&&\frac{\partial L}{\partial \boldsymbol{v}_{l}} = \sum_{m} 2 \left ( H_{\rm FM} (\{ x_{i}^{m} \}) - H^{m} \right ) \frac{\partial}{\partial \boldsymbol{v}_{l}} H_{\rm FM}(\{ x_{i}^{m} \}) \nonumber \\
&&= \sum_{m} 2 \left ( H_{\rm FM} (\{ x_{i}^{m} \}) - H^{m} \right ) \frac{\partial}{\partial \boldsymbol{v}_{l}} 
\sum_{i,j=1}^{N} \langle \boldsymbol{v}_{i},\boldsymbol{v}_{j} \rangle x_{i}^{m} x_{j}^{m} \nonumber \\
&&= \sum_{m} 2 \left ( H_{\rm FM} (\{ x_{i}^{m} \}) - H^{m} \right ) \left ( \sum_{i \neq l}^{N} \boldsymbol{v}_{i} x_{i}^{m} \right ) x_{l}^{m},
\label{eq:gradient_v}
\end{eqnarray}
where $l$ denotes any identification number of the binary variable and parameters as well as $i$.
From Eqs.~\ref{eq:gradient_b} and~\ref{eq:gradient_v}, 
if the value of $x_{l}^{m}$ is zero for all $m$, then the gradients are zero.
In other words, the gradients are zero if the dataset does not contain the pair of evaluation point and evaluation value where $x_{l} = 1$.
The zero gradients mean that the parameters $b_{l}$ and $\boldsymbol{v}_{l}$ are not updated at all.
Without parameter updates, the random noise-like initial values of $H_{\rm FM}$ remains.
This is the origin of the noisy function surface of $H_{\rm FM}$.
Note that reducing the expressiveness of the model by lowering the rank of the FM may prevent overfitting, but it does not improve the zero gradients.
This is because the rank of the FM is unrelated to the update of the parameters.

The above problem motivated the development of a method to overcome it.
The loss function should be modified so that the parameters are updated even when there are no samples with $x_{l}=1$ in the dataset.
Specifically, we introduce a function smoothing regularization (FSR) for the loss function expressed as
\begin{eqnarray}
L' = L + \lambda_{\rm SR} \left [ \sum_{(p,q) \in \mathcal{A}} || \boldsymbol{v}_{p} - \boldsymbol{v}_{q} ||^{2} + (b_{p} - b_{q})^{2} \right ],
\label{eq:function_smoothing}
\end{eqnarray}
where $\lambda_{\rm SR}$ is the strength of the smoothing regularization, and $\mathcal{A}$ is the set of all adjacent pairs of binary variables representing continuous values.
Specifically, $\mathcal{A} = \{ (1,2),(2,3),... ,(C-1,C),(C+1,C+2),... (2C-1,2C) \}$.
Note that $(C,C+1)$ is excluded from adjacent pairs because $C$ and $C+1$ belong to different continuous values.

Figure~\ref{fig:h1_compare}(e), (f), (g), and (h) show the function surfaces obtained by function smoothing regularization FM (FSRFM) with $\lambda_{\rm SR}=0.1$ immediately after steps 1, 2, 8, and 16, respectively.
In contrast to naive FM, FSRFM approached the true Hamiltonian as the size of the dataset increased with the number of steps.
Even regions with sparse evaluation points were effectively sampled after the first step.
We demonstrated that FSRFM substantially progresses the black-box optimization and allows for efficient search for evaluation points.

\begin{figure}
\includegraphics[width=0.48\textwidth]{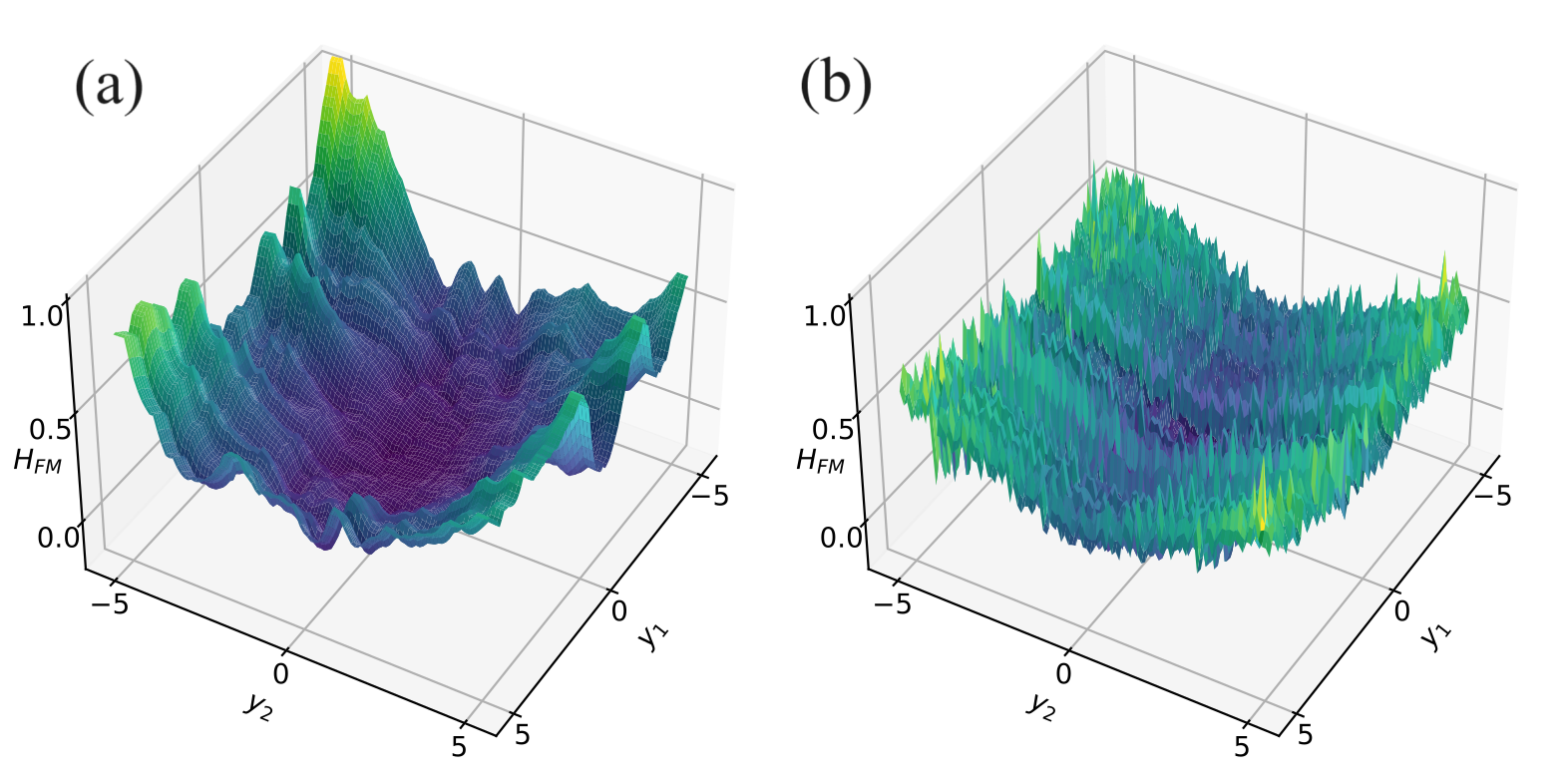}
\caption{
Function surfaces obtained by minimizing the loss functions (a) $L'$ and (b) $L''$ over 16 steps.
}
\label{fig:compare_l2_fs}
\end{figure}

Let us explain the characteristics of FSRFM.
The minimization of the second term on the right side of Eq.~\ref{eq:function_smoothing} (the smoothing regularization term) has the effect of bringing the values of adjacent parameter pairs ($\{ b_{p},b_{q} \}$ and $\{ \boldsymbol{v}_{p},\boldsymbol{v}_{q} \}$) closer together.
This not only stabilizes the parameter values, but also smoothes the surface of the $H_{\rm FM}$.
When the number of steps is small, i.e., the dataset is relatively small, problems based on the random-like function surface become more pronounced, making the smooth function surface even more important.
FSR is distinctly different from the commonly used L2 regularization\cite{luo2016regression}.
The loss function $L''$ with L2 regularization introduced is as the following equation,
\begin{eqnarray}
L'' = L + \lambda_{\rm L2} \left[ \sum_{p} || \boldsymbol{v}_{p} ||^{2} + (b_{p})^{2} \right ],
\label{eq:l2}
\end{eqnarray}
where $\lambda_{\rm L2}$ is the strength of L2 regularization.
Figure~\ref{fig:compare_l2_fs}(a) and (b) show the resulting function surfaces obtained by minimizing the loss functions $L'$ and $L''$ over 16 steps, respectively.
The L2 regularization produced a slight improvement over the naive FM (Fig.~\ref{fig:h1_compare}(a)-(d)), but the noise remained.
This is because L2 regularization simply applies a constraint on the norm of parameter values.
This restriction guarantees initial value updates, but does not guarantee smoothness of $H_{\rm FM}$.
FSR, on the other hand, estimated $H_1$ well even with a small dataset (Fig.~\ref{fig:h1_compare}(e)-(h)).
This is because the FSR guarantees both the update of initial values and the smoothness of $H_{\rm FM}$.

\subsection{Generalization performance of the function smoothing regularization}
Next, we evaluated the generalization performance of FSRFM by applying it to the interactive toy Hamiltonian $H_2$.
Note that the generalization performance can be evaluated only from the FM operation in Fig.~\ref{fig:flowchart}.
The learning rate of AMSGRAD was set to 0.1.
The dataset were sampled uniformly and randomly.
The number of samples $N_{\rm s}$ was set to 10, 50, 100, 500, and 1000, $\lambda_{\rm SR}$ was set to 0, 0.1, 1, and 10, and the rank of FM ($K$) was set to 4 and 16 for a total of 30 conditions were tried.
Note that $\lambda_{\rm SR}=0$ implies naive FM and that the same dataset was used for the same $N_{\rm s}$ conditions.

\begin{figure}
\includegraphics[width=0.48\textwidth]{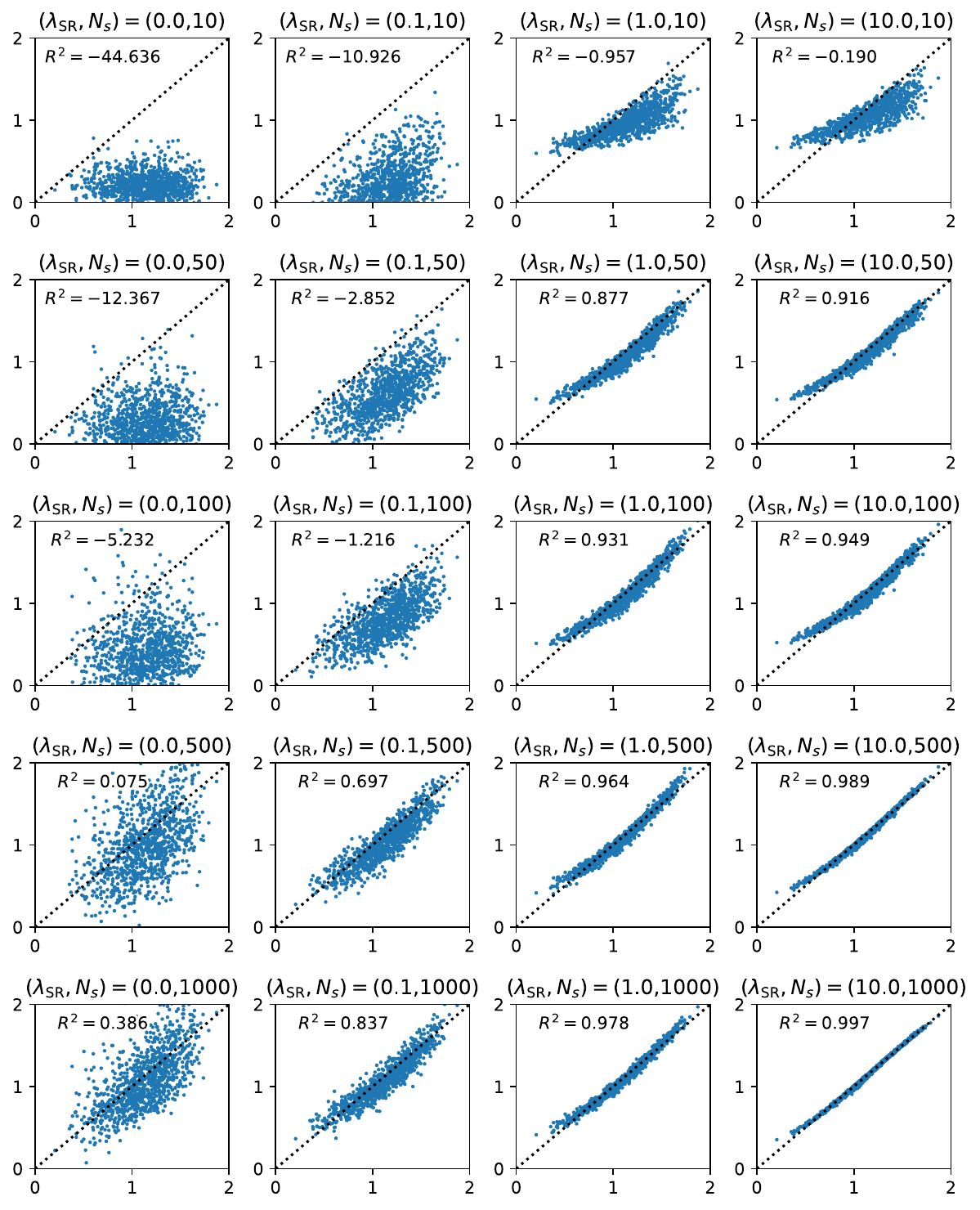}
\caption{
$H_2$-$H_{\rm FM}$ plots of test dataset for FM rank of 16.
}
\label{fig:h2_rank16}
\end{figure}

\begin{figure}[b]
\includegraphics[width=0.48\textwidth]{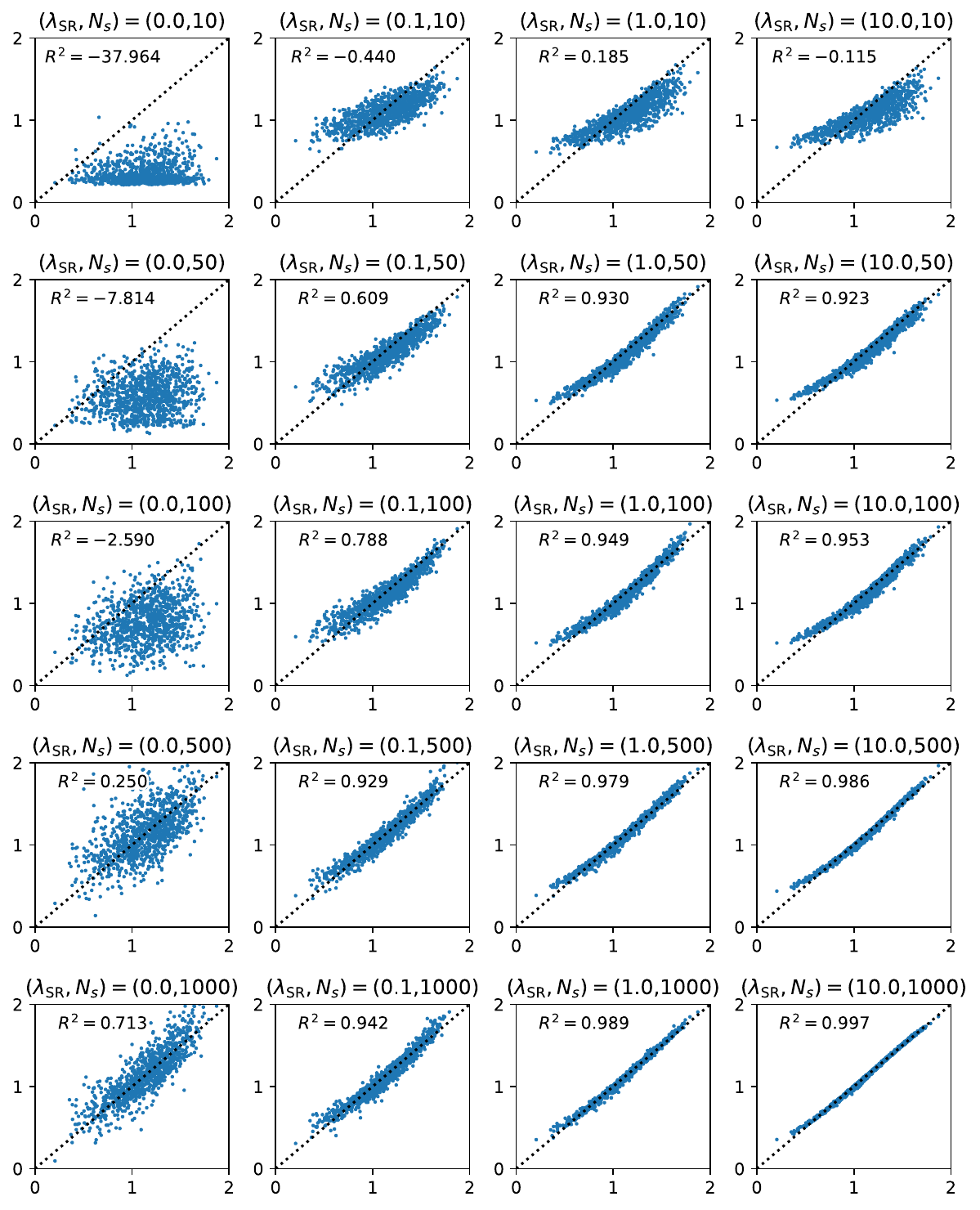}
\caption{
$H_2$-$H_{\rm FM}$ plot of test dataset for FM rank of 4.
}
\label{fig:h2_rank4}
\end{figure}

Figures~\ref{fig:h2_rank16} and ~\ref{fig:h2_rank4} show $H_2$-$H_{\rm FM}$ plots of test dataset for FM ranks of 16 and 4, respectively.
The closer the plots are to the straight line with slope 1, the better the performance of the FMs in estimating $H_2$, but more quantitatively with the coefficient of determination expressed as the following equation,
\begin{eqnarray}
R_{2} = 1 - \frac{\sum_{m} \left ( H_{\rm FM}^{m} -  H^{m} \right )^{2}}{\sum_{m} \left ( H_{\rm FM}^{m} - \overline{H_{\rm FM}} \right )^2},
\label{eq:determ_coeff}
\end{eqnarray}
where $\overline{H_{\rm FM}}$ is the mean value of $H_{\rm FM}^{m}$.
The maximum value of the coefficient of determination is clearly 1.
From Fig.~\ref{fig:h2_rank16}, it is clear that the FSRFM ($\lambda_{\rm SR} > 0$) shows higher generalization performance than naive FM ($\lambda_{\rm SR}=0$).
It is worth noting that the coefficient of determination increases as $\lambda_{\rm SR}$ increases.
At $\lambda_{\rm SR}=10$, the FSRFM achieves a coefficient of determination above 0.9 with only 10 samples.
In contrast to the FSRFM, naive FM ($\lambda_{\rm SR}=0$) had a generally low coefficient of determination.
Naive FM required 500 samples to achieve even a zero coefficient of determination.
Note that a zero coefficient of determination corresponds to estimating the mean of the true Hamiltonian.

Figure~\ref{fig:h2_rank4} shows the results of reducing the ranks of FM compared to Fig.~\ref{fig:h2_rank16}.
Since the reduction in ranks is a reduction in the expressiveness of the model, generalization performance should improve.
In other words, the conditions are relatively favorable for naive FMs.
Nevertheless, the naive FM improved only slightly.
For example, at $N_{\rm s}=100$, the coefficient of determination for naive FM improved from -5 (see Fig.~\ref{fig:h2_rank16}) to -2 with the rank reduction, but the FSRFM at $\lambda_{\rm SR}=10$ outperformed it even $N_{\rm s}=10$.
As noted above, rank reduction is a reduction in expressive capacity and a straightforward way to improve generalization performance.
FSRFM, on the other hand, maintained expressive power and achieved high generalization performance with the bias from regularization.
Furthermore, since rank reduction also improved the generalization performance of FSRFM, there is significant merit in performing FSR regardless of rank.

\subsection{Prediction of nanophysical properties of contrast agents}
Finally, a practical problem was solved using the QA with FSRFM (FSRFMQA).
That is, the FSRFM was used to train the Hamiltonian $H_3$ and QA was performed to estimate the nanophysical parameters of the contrast agent.
The initial samples of the dataset were chosen uniformly and randomly.
The step shown in Fig.~\ref{fig:flowchart} was repeated 16 times with the following conditions: the learning rate of AMSGRAD was set to 0.1 and 1000 gradient updates were performed, the rank of FM was set to 16, and $Q_{ij}$ was input to Dwave's Simulated Annealing Sampler to obtain 16 new evaluation points.
Since FMQA is a heuristic, 16 independent trials were performed to obtain statistics.
For comparison, the above was also done for QA with naive FM (naive FMQA).

\begin{figure}
\includegraphics[width=0.48\textwidth]{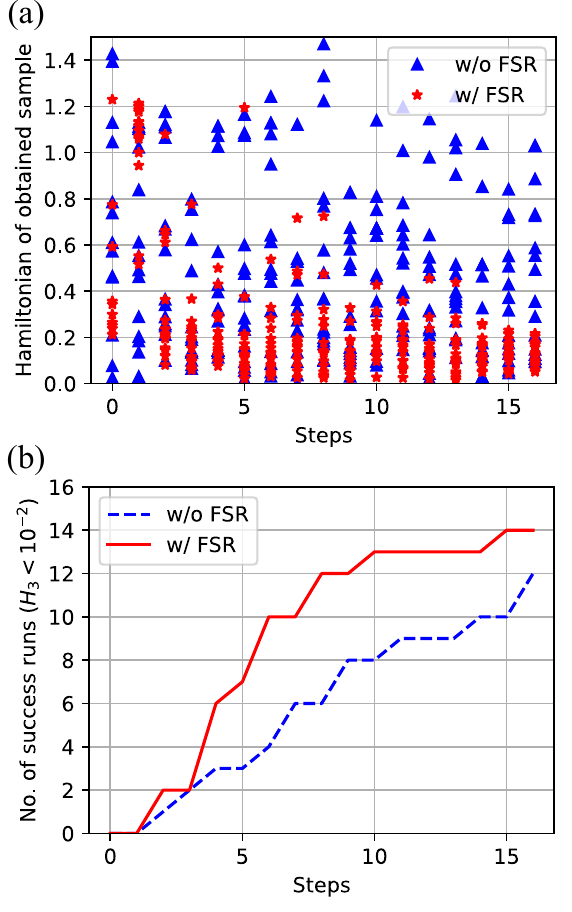}
\caption{
Dependencies on the number of steps of (a) the evaluation values corresponding to the obtained evaluation points, and of (b) the number of trials where the evaluation value is within 0.01 of the minimum value of $H_3$.
}
\label{fig:h3}
\end{figure}

Figure~\ref{fig:h3} (a) shows the dependence of the evaluation values corresponding to the obtained evaluation points on the number of steps.
Note that the evaluation values were determined by substituting the obtained evaluation points into the true Hamiltonian equation (see Eq.~\ref{fig:h3}).
The results of one out of 16 independent trials are plotted.
Note that 16 evaluation points and values are generated per step, so 16 evaluation values are plotted per step.
The evaluation values corresponding to the initial evaluation points are plotted as 0 steps.
The FSRFMQA obtained evaluation points that yielded lower evaluation values as the number of steps increased.
This implies that the training of FSRFM is improved by increasing the size of the dataset, and regions with smaller true Hamiltonians ($H_3$) are explored more efficiently.
In contrast, naive FMQA did not result in lower evaluation values as the number of steps increased.
As reiterated in previous sections, this is because the function surface of $H_{\rm FM}$ is noisy.

Figure~\ref{fig:h3} (b) shows the dependence on the number of steps of the number of trials where the evaluation value is within 0.01 of the minimum value of $H_3$.
FSRFMQA was able to search for evaluation values closer to the minimum in a shorter number of steps than naive FMQA.
FSRFMQA obtained the evaluation values of the same quality in roughly half the number of steps of naive FMQA.
These results demonstrate that the FSR is beneficial in the practical parameter estimation problem.

\section{Conclusions}
In this study, we investigated in detail the function surfaces of Hamiltonians obtained by FM in the widely common case where real numbers were represented by a combination of binary variables.
The function surface of the simple toy Hamiltonian $H_1$ obtained from naive FM was very noisy, implying that it is difficult to propose the next evaluation point from this function surface.
In fact, the sampling of new evaluation points was extremely biased and the black box optimization did not work.
The reason why the naive FM obtains a very noisy function surface is due to the minimization process of the loss function.
If the dataset does not contain a pair of evaluation point and evaluation value corresponding to the FM parameters, the gradients that prompt the parameters to be updated is zero, and the parameters are not updated.
The fact that the FM parameters are not updated means that the estimated Hamiltonian will continue to have random noise-like initial values.
This is the origin of the noisy functional surfaces that naive FMs acquire.

The above problem motivated the development of a method to overcome it.
By introducing FSR into the loss function, the FM parameters were updated even when the dataset did not contain the pair of evaluation point and evaluation value corresponding to the parameters.
The Hamiltonian obtained by FSRFM approached the true Hamiltonian as the number of steps increased.
This is because FSR not only effectively updates the FM parameters, but also smoothes the function surface of Hamiltonian.
We demonstrated that FSRFM significantly advances black-box optimization and enables efficient search for evaluation points.
The generalization performance of FSRFM was then evaluated with the interactive toy Hamiltonian $H_2$.
The performance of FSRFM was generally better than that of naive FM.
It is worth noting that the coefficient of determination increased as $\lambda_{\rm SR}$ increased.
FSRFM achieved the coefficient of determination above 0.9 with only 10 samples when $\lambda_{\rm SR} = 10$.
Even when the expressiveness of the FM model was reduced, FSRFM still demonstrated higher generalization performance than naive FM.
Finally, the FSRFM was used to train the Hamiltonian $H_3$ and QA was used to estimate the nanophysical parameters of the contrast agent.
The training of FSRFM was greatly improved by increasing the size of the dataset, and the region with small true Hamiltonian ($H_3$) were efficiently explored.
FSRFMQA was able to search for near-minimal values in the shorter number of steps (approximately half) than naive FMQA.
We demonstrated that FSR is useful for practical property estimation problems.

From the above, FSR is a simple and general method that should be introduced into FM in widely common cases where integers and real numbers are represented by a combination of binary variables.
FSR overcomes the weaknesses of naive FM and preserves the inherent capabilities of QA, thereby contributing to solving problems that were previously considered unsolvable by FMQA.
It is expected that FSRFMQA will solve a wider variety of continuous variable optimization problems than ever before.

\begin{acknowledgments}
This paper is based on results obtained from a project, JPNP23003, commissioned by the New Energy and Industrial Technology Development Organization (NEDO).
\end{acknowledgments}



\nocite{*}

\bibliography{apssamp}

\end{document}